\documentclass[twocolumns]{aa}

\usepackage{graphicx}
\begin{document}
\title{The HELLAS2XMM survey: VI. X-ray absorption in the 1df
AGN sample through a spectral analysis.
\footnote{Based on observations made with XMM-Newton,
an ESA science mission.}}

\author{G.C. Perola\inst{1,2}, S. Puccetti\inst{1,3}, F. Fiore\inst{1}, 
N. Sacchi\inst{2}, M. Brusa\inst{4,5}, F. Cocchia\inst{1,3},
A. Baldi\inst{6}, N. Carangelo\inst{7}, P. Ciliegi\inst{5}, 
A. Comastri\inst{5}, F. La Franca\inst{2}, R. Maiolino\inst{8}, 
G. Matt\inst{2}, M. Mignoli\inst{5}, S. Molendi\inst{6}, 
C. Vignali\inst{5}
}
\offprints{G. C. Perola, perola@fis.uniroma3.it}

\institute{INAF, Osservatorio Astronomico di Roma, via Frascati 33, 
Monteporzio-Catone (RM), 00040 Italy
\and 
Dip. di Fisica, Universit\`a Roma Tre
\and
Dip. di Fisica Universit\`a di Roma Tor Vergata
\and
Dip. di Astronomia Universit\`a di Bologna
\and
INAF-Osservatorio Astronomico di Bologna
\and
Harvard-Smithsonian Center for Astrophysics
\and
IASF/CNR Milano
\and
INAF-Osservatorio Astrofisico di Arcetri
}

\date{January 22,  2004}

\abstract{The spectroscopic analysis of 117 serendipitous sources in
the HELLAS2XMM 1df (1 degree field) survey is described. Of these,
106 sources, of which 86\% have a spectroscopic redshift, are used to evaluate the
fraction of X-ray absorbed (log N$_H$$>$22) Active Galactic Nuclei
(AGN) in the 2--10 keV flux range 0.8--20$\times$10$^{-14}$ erg
cm$^{-2}$ s$^{-1}$. This fraction turns out lower than what is predicted by
two well known Cosmic X-Ray Background synthesis models, and the
discrepancy is significant at the 99.999\% level. This result
consolidates the findings recently obtained by other authors. In the
flux interval explored, the data are consistent with an intrinsic
distribution of the absorbing columns (flat per decade above
log$N_H>$21) independent of luminosity and redshift, together with an
AGN luminosity function evolving purely in luminosity. It is shown
that, on the other hand, extrapolation to lower fluxes fails to
reproduce the results inferred from the Chandra Deep Field North
survey. It is found that about 40\% of the high luminosity
sources in our sample have best fit logN$_H$$>$22, and the surface
density of these X--ray obscured QSOs can then be estimated at about 48 per
square degree, at the flux limit of $\sim10^{-14}$ erg cm$^{-2}$ s$^{-1}$ of the
HELLAS2XMM 1df survey.  As a side issue, 5 or 6 out of 60 sources,
that is about 10\%, identified with broad line AGN, turn out to be
affected by logN$_H$$>$22 absorption.


}

\authorrunning {Perola et al.}
\titlerunning {HELLAS2XMM, X-ray absorption in AGN sample}

\maketitle

\section{Introduction}

After the success of ROSAT (Hasinger et al. 1998) in resolving
about 75\% of the X-ray background (XRB) in the 0.5--2 keV band
into sources largely associated with Active Galactic Nuclei (AGN),
the satellites Chandra and XMM-Newton achieved a similar result, 
up to at least 85\% of the XRB, in the 2-10 keV band (Mushotzky et al. 2000,
Giacconi et al. 2001, 2002, Hasinger et al. 2001; 
Alexander et al. 2003; see also Moretti et
al. 2003 and references therein). The combination of the results
in the two bands provides also the observational support
for the intuition by Setti \& Woltjer (1989) that the XRB 
could be explained by a dominant contribution of AGN,
affected by photoelectric obscuration in different proportions
over a wide range of gas columns N$_H$. This suggestion
led to several attempts, all formally successful,
to synthesize the XRB starting from somewhat different
assumptions about the AGN Luminosity Function (LF) and
its cosmological evolution, and N$_H$ distributions
(e.g. Comastri et al. 1995, Gilli et al. 2001, Wilman \& Fabian 1999,
Miyaji et al. 2000, Ueda et al. 2003). In this context, an important
issue, which is being explored with increasingly
more detailed X--ray spectral analysis 
and spectroscopic identification of the
optical counterparts, is the fraction of sources
with different intrinsic N$_H$ as a function
of their flux. This approach provides very
strong constraints, especially when accompanied
by the study of the LF performed using the same 
data (e.g. Ueda et al. 2003). The present limits
are set by progressively poorer
statistics in the X-ray spectra and in the
optical spectroscopic identification as one
goes to fainter sources. Thus, while a treatment
as just outlined, based on the full ensemble
of sources utilized by Fiore et al. (2003, hereafter Paper IV),
is deferred to La Franca et al. (in prep.),
this paper aims to exclusively present
the information on the fraction of sources
affected by different levels of X-ray obscuration,
down to a limit in F(2-10 keV) of about 10$^{-14}$ 
erg cm$^{-2}$ s$^{-1}$
(corresponding to about 35\% of the XRB), extracted from
the HELLAS2XMM 1df sample. This sample comprises
117 sources, 93 of them (80\%) with a spectroscopic
redshift available.     
 
The spectral counts extraction is described in Sect. 2,
their best fit analysis in Sect. 3, the synthesis of
the results in Sect. 4. Sect. 5 is devoted to a discussion
of the results compared to XRB synthesis models, Sect. 6 
to the conclusions. 

\section{Extraction of the spectral counts}

The HELLAS2XMM 1df (1 degree field) sample is composed of 122
sources (Paper IV), serendipitously detected in the 2-10 keV band
in five XMM-Newton fields: PKS 0537-286, PKS
0312-770, A2690, G158-100, Mrk 509 (see Baldi et al. 2002, 
for the epochs and exposure times). The
observations were performed with the European Photon
Imaging Camera (EPIC), composed by one pn back-illuminated CCD
array (Str\"{u}der et al. 2001) and by two mos
front-illuminated CCD arrays (Turner et al. 2001), named mos1 and
mos2 respectively. However only the fields PKS 0537-286, PKS 0312-770, A2690
have been observed with the three cameras, whereas the
G158-100, Mrk 509 fields have been observed with mos1 and mos2
alone.

The source counts in each camera have been obtained using the events
files, in the energy range 0.5-10 keV for the pn and 0.3-10 keV for
the mos. The counts of the two mos cameras were eventually combined.
The counts of each source have been extracted in a circular region
with a radius in the range $20\arcsec$--$40\arcsec$. In general the
radius value was chosen so that the S/N ratio was roughly optimized,
but in a few cases this choice was limited by the presence of nearby
sources, or by a peculiar position of the source on the detector, for
example close to a gap in the CCD array. In some cases the source was
detected, and the corresponding counts extracted, only in either the
pn or in one or both of the mos cameras, because pn and mos do not cover
exactly the same sky regions, and the position of the gaps differs in
the pn, mos1 and mos2 CCD arrays.

The background counts for each source were extracted from the
nearest source-free region. In doing so, areas near gaps in the 
CCD array and near the edge of the telescope 
field of view have been excluded, as well as regions containing hot pixels
and other CCD cosmetic defects.
 
The ancillary response files were generated for each source by means
of the tool {\sc arfgen} (SAS 5.4.1\footnote{
http://xmm.vilspa.esa.es/external/xmm\_sw\_cal/\\
sas\_frame.shtml}), in order to
properly correct for energy dependent vignetting and point spread
function. The response matrix file, updated for all the observation
modes to January 29, 2003 and available at the XMM-Newton
archive\footnote{
ftp://xmm.vilspa.esa.es/pub/ccf/constituents/extras/\\responses/}, was
adopted.

Among the 122 sources in Paper IV, the one identified with a star (05370006)
was discarded from the start. For the two extended sources, 03120008 and
26900013, the spectral analysis revealed the presence of
an AGN contribution in the first, which was therefore kept in the sample. Moreover,
only pn and mos spectra with combined counts greater than 40 were
considered, and the sources 05370159, 05370164 and 03120116 were
therefore discarded.

In summary, the sample studied in this paper is composed of 117 sources,
with a spectroscopic redshift available for 93 of them, as reported
in Paper IV.

\section{Spectral fits}

The spectral counts, when higher than about 120, were first
accumulated in energy bins with 20 counts each, from 0.3 keV
to 10 keV in the mos, and from 0.5 to 10 keV in the pn. They were then fitted,
using XSPEC (version 11.2.0) and the $\chi^2$ statistic, with the
simple model comprising, in addition to the known galactic absorption:
(1) a power law, with two parameters, normalization and photon spectral
index ${\Gamma}$; (2) the absorption N$_H$ at the redshift of the optical
counterpart; when both pn and mos data were available, their relative
normalization mos/pn was left free to vary between 0.8 and 1.2. This
interval was chosen conservatively wider than applicable on-axis, 
because for sources off-axis a fully reliable
intercalibration is still lacking. When the spectral counts were lower
than about 120, the C statistic (Cash 1979) was used instead, as
implemented in XSPEC (Arnaud 2003 \footnote{K.A. Arnaud, 2003, "XSPEC
User Guide for version 11.3"\\
http://heasarc.gsfc.nasa.gov/docs/software/lheasoft/xanadu/\\
xspec/manual/manual.html}) 
after background subtraction (see Alexander et al. 2003a for a
similar procedure) and with 5 counts in each energy bin (the latter
choice was made only for convenience; it does not impair the correct
use of the embedded statistics when using the abovementioned XSPEC
implementation). In this case
the normalization mos/pn was set equal to 1.
  
The systematic use of the simple model is meant to yield an
'effective' value for the absorbing column, the best one can obtain
with the relatively modest statistics available. In addition,
it should be stressed that this is after all the most meaningful
quantity for the implications that absorption has on the 
synthesis of the X-ray background.
 
The galactic absorption columns adopted (see Baldi et al. 2002) are:
8$\times$10$^{20}$ cm$^{-2}$ for the field PKS0312, 2$\times$10$^{20}$
cm$^{-2}$ for A 2690, 2.1$\times$10$^{20}$ cm$^{-2}$ for PKS0537,
4$\times$10$^{20}$ cm$^{-2}$ for Mkn 509, 2.5$\times$10$^{20}$
cm$^{-2}$ for G158--100.

The sources with a spectroscopic redshift and those without were
treated separately, and the whole set was subdivided into five
subsets.  The first subset (S1) comprises spectra of objects with
known z, whose number of degrees of freedom (dof) is equal to or
larger than 8 when both ${\Gamma}$ and N$_H$ are left as free
parameters, and the mos/pn normalization is frozen to its best fit
value before estimating the errors. This corresponds to a
total number of counts equal to or larger than 220. The 90{\%} confidence
intervals on N$_H$ and ${\Gamma}$ were therefore calculated with
$\Delta\chi^2$ = 4.61.  The results for the 44 S1 sources are
presented in Table 1, where the following information is given: in
Col. 2 the optical classification (AGN1 and AGN2 with their usual
meaning, but see Paper IV, ELG = Emission Line Galaxy, ETG = Early Type
Galaxy; the few objects reclassified differently from Paper IV are starred), in
Col. 3 the redshift, in Col. 4 the instrument(s) used (pm = pn and mos
combined), in Col.s 5, 6 and 7, N$_H$, ${\Gamma}$ and $\chi^2$/dof, in
Col. 8 the 2-10 keV flux F as observed (when applicable, the mean of
the pn and mos values), in Col. 9 the same corrected for the
absorption F$_u$, and the corresponding luminosity in Col. 10,
computed using a cosmology with H$_0$ = 70 km s$^{-1}$ Mpc$^{-1}$,
${\Omega}_m$ = 0.3 and ${\Lambda}$ = 0.7, and the K-correction
appropriate for the best fit value of ${\Gamma}$. Note that throughout
the next sections F$_u$ will be used.

\begin{table*}[e]
\begin{center}
\caption{\bf Spectral fits of the subset S1}
\begin{tabular}{l|c|c|c|c|c|c|c|c|c}
\hline
source ID &  type  &  z   &   inst&  N$_H$$^a$ & $\Gamma$ &  $\chi^2$/dof
&  F$^b$ &  F$_u$$^b$ & logL$_{2-10keV}$$^c$\\

\hline
03120002 &  AGN1  &    1.187  &   pm     & $<$0.05 &
1.88$\pm_{0.06}^{0.07}$ & 190.0/183 & 41.0 & 41.3 & 45.48\\
03120003 &  AGN1   &   0.420  &   pm   & $<$0.03 &
1.98$\pm_{0.11}^{0.12}$ & 88.4/72 & 14.8 & 14.9 & 43.97 \\
03120004  & AGN1 &      0.890  &   pm   & $<$0.08 & 2.44$\pm_{0.15}^{0.16}$ & 62.6/53 &
5.2 & 5.2  & 44.43\\
03120005 &  AGN1 &      1.274 &         pm &$<$0.14 &
1.83$\pm_{0.15}^{0.16}$ & 43.3/46 & 9.4 & 9.4 & 44.89\\
03120006 &  AGN2 &      0.680 &         pm  &$<$0.16 &
1.56$\pm_{0.14}^{0.17}$ & 33.1/39 & 10.8 & 10.8 & 44.24\\
03120007 &  AGN1 &      0.381 &         m    & 0.12$\pm_{0.12}^{0.24}$
& 1.51$\pm_{0.35}^{0.41}$ & 17.0/18 & 19.5 & 19.7 & 43.93\\
03120009  & AGN1 &      1.522 &         pm  & $<$0.84 &
1.86$\pm_{0.33}^{0.49}$ & 16.4/19 & 2.4 & 2.4 & 44.49 \\
03120010  & AGN1 &      0.246 &         pm   & $<$0.14 &
2.69$\pm_{0.48}^{0.79}$ & 24.1/19 & 1.3 & 1.3 & 42.44\\
03120012 &  AGN1 &      0.507 &         pm   &$<$0.24 &
2.49$\pm_{0.44}^{0.79}$ & 20.0/16 & 2.1 & 2.2  & 43.43\\
03120013 &  AGN1 &      1.446 &         pm   & 0.71$\pm_{0.71}^{2.10}$
& 2.83$\pm_{0.80}^{1.51}$ & 18.4/14 & 0.8 & 0.8  & 44.34\\
03120014 &  ELG &       0.206 &         pm   & $<$0.26 & 1.54$\pm_{0.31}^{0.53}$ &
21.5/18 & 5.3 & 5.3  & 42.78\\
03120017 &  ETG  & 0.320 &      pm    & 0.13$\pm_{0.13}^{0.45}$ &
2.28$\pm_{0.66}^{1.15}$ & 14.1/11 & 1.9 & 2.0 & 42.86\\
03120018  & ETG  & 0.159 &      pm    & 0.46$\pm_{0.46}^{0.84}$ &
1.87$\pm_{0.78}^{1.08}$ & 22.4/14 & 2.4 & 2.5  & 42.23\\
03120020  & ELG  & 0.964 &      pm    &$<$0.70 &
2.03$\pm_{0.41}^{0.70}$ & 9.5/12 & 2.4 & 2.4  & 44.07\\
03120021 &  AGN1  &   2.736 &   pm   &  $<$4.19 &
1.53$\pm_{0.52}^{0.69}$ & 10.4/11 & 2.1 & 2.2  & 44.87\\
03120028  & ELG  & 0.641 &      pm   & 0.54$\pm_{0.54}^{2.70}$ &
1.63$\pm_{0.80}^{1.54}$ & 9.8/12 & 2.2 & 2.2 & 43.51\\
26900001  &  AGN1 & 0.336 &     pm   & $<$0.06 &
1.78$\pm_{0.14}^{0.20}$ & 42.8/43 & 8.4 & 8.4 & 43.47\\
26900002  &  AGN1 &     0.850 &         pm  & 0.54$\pm_{0.30}^{0.29}$
& 1.63$\pm_{0.24}^{0.28}$ & 32.4/30 & 14.6 & 14.8  & 44.62\\
26900003  &  AGN1 &     0.433 &         pm  & $<$0.08 &
2.13$\pm_{0.20}^{0.29}$ & 33.1/34 & 6.7 & 6.7 & 43.68\\
26900004  &  AGN1 &     0.284 &         pm  & $<$0.05 &
2.03$\pm_{0.21}^{0.26}$ & 30.4/26 & 7.8 & 7.8 & 43.3 \\
26900007  &  AGN1 &     1.234 &         pm  & $<$0.29 &
2.07$\pm_{0.32}^{0.43}$ & 14.9/12 & 2.1 & 2.1 & 44.29\\
26900010   & AGN1 &     1.355 &         pm  & $<$0.38 &
1.91$\pm_{0.37}^{0.47}$ & 8.5/10 & 2.9 & 2.9  & 44.47\\
26900012  &  AGN1 &     0.433 &         pm  & 0.14$\pm_{0.14}^{0.27}$
& 2.66$\pm_{0.69}^{1.09}$& 7.4/15 & 0.8 & 0.8 & 42.85\\
26900015  &  AGN1 &     1.610 &         pm & 1.52$\pm_{1.27}^{2.47}$ &
2.72$\pm_{0.74}^{1.28}$ & 8.8/14 & 1.1 & 1.1 & 44.57\\
05370002  &  AGN1  &   1.244 &  pm  & $<$0.10 & 1.95$\pm_{0.09}^{0.10}$
& 85.0/96 & 15.5 & 15.5 & 45.13\\
05370003  &  AGN1  &   0.317 &  pm   & $<$0.10 &
2.04$\pm_{0.19}^{0.23}$ & 37.1/52 & 10.1 & 10.1 & 43.53\\
05370004  &  AGN1 &    0.894 &  pm   &  $<$0.24 &
1.58$\pm_{0.14}^{0.20}$ & 55.9/47 & 8.1 & 8.1 & 44.39\\
05370005  &  AGN1  &  1.158 &   pm   & 0.16$\pm_{0.16}^{0.69}$ &
1.60$\pm_{0.34}^{0.42}$ & 15.7/14 & 11.1 & 11.1  & 44.79\\
05370007  &  AGN1  &  0.842 &   pm   & $<$0.14 &
1.91$\pm_{0.25}^{0.34}$ & 20.9/21 & 2.7 & 2.7  & 43.94\\
05370008  & AGN2$^{*}$ &  0.379 &   pm   & $<$0.18 &
2.29$\pm_{0.33}^{0.48}$ & 17.8/25 & 3.9 & 4.0 & 43.34\\
05370009  &  AGN1 &   0.770 &   pm   &  0.14$\pm^{ 1.20 }_{0.14}$  &
2.10$\pm^{ 1.28}_{ 0.53}$ & 16.8/16 & 2.2 & 2.2  & 43.81\\
05370013   & AGN1 &   0.901 &   pm   & $<$0.39   & 1.85$\pm^{ 0.53}_{
  0.35}$  & 11.5/11 & 3.0 & 3.0  & 44.04\\
05370014  &  AGN1  &  1.659 &   pm   & $<$1.32 &
1.25$\pm_{0.35}^{0.54}$ & 7.7/14 & 6.3 & 6.3 & 44.74\\
05370015  &  AGN1  &  0.880 &   pm   & 0.37$\pm_{0.37}^{1.17}$ &
2.32$\pm_{0.78}^{1.53}$ & 11.6/13 & 2.4 & 2.5 & 44.07\\
05370016   & AGN2  &  0.995 &   pm  &  1.32$\pm_{0.88}^{1.56}$ &
2.05$\pm_{0.52}^{0.81}$ & 9.8/17 & 3.5 & 3.6 & 44.28\\
05370017  &  AGN1  &  0.904 &   pm  & $<$0.28 &
1.86$\pm_{0.36}^{0.54}$ & 6.4/9 & 2.7 & 2.7 & 44.0\\
05370021  & ELG$^{*}$ &  1.192 &   pm  &  0.16$\pm_{0.16}^{1.50}$ &
1.54$\pm_{0.54}^{0.99}$ & 9.0/11 & 3.9 & 3.9 & 44.34\\
05370024  &  ETG   &  0.075 &   pm  &  $<$0.12 &
1.18$\pm_{0.36}^{0.42}$  & 19.0/12 & 4.0 & 4.0 & 41.72\\
0537011a  &  AGN2  &  0.981 &   pm  & 1.33$\pm_{0.90}^{1.50}$ &
1.88$\pm_{0.40}^{0.63}$ & 26.7/16 & 4.1 & 4.2 & 44.29\\
50900020  &  AGN1  &  0.770  &  m   &  0.48$\pm_{0.48}^{1.53}$ &
2.42$\pm_{0.87}^{1.51}$ & 5.1/8 & 4.2 & 4.3 & 44.18\\
50900031  &  AGN1 &     0.556  &  m  & $<$0.10 &
1.83$\pm_{0.64}^{0.67}$ & 21.2/11 & 4.3 & 4.3 & 43.69\\
15800001  &  AGN1   & 1.211  &  m    & $<$0.31&
2.12$\pm_{0.22}^{0.36}$ & 14.8/19 & 8.7 & 8.7  & 44.91\\
15800002  &  AGN1 &     0.848 &   m  &   $<$0.13 & 1.94$\pm^{ 0.30}_{
  0.26 }$ & 17.0/17 & 7.3 & 7.3   & 44.39\\
15800008  &   AGN1 &    1.151  &   m  & $<$1.42 &
1.43$\pm_{0.44}^{0.76}$ & 6.2/8 & 3.9 & 3.9 & 44.27\\

\hline
\end{tabular}
\end{center}
$^a$ N$_H$ in source frame, units of 10$^{22}$ cm$^{-2}$;
$^b$ Flux in the 2--10 keV band, F: observed, F$_u$: corrected
for absorption, in units of 10$^{-14}$ erg cm$^{-2}$ s$^{-1}$;
$^c$ Log of the luminosity in units of erg s$^{-1}$.
\end{table*}

\begin{figure}
\centering
\includegraphics[angle=-90,width=9cm]{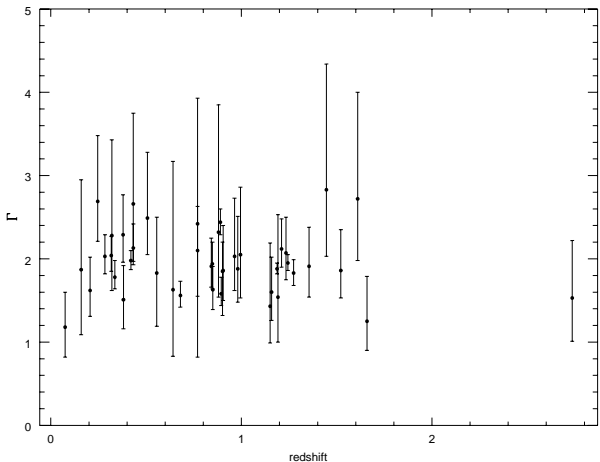}
\caption{The spectral index $\Gamma$ of the S1 sources as a function
of redshift.
}
\label{FigGam}
\end{figure}

The distribution of $\Gamma$ from Table 1 is given in Fig.
\ref{FigGam} as a function of the redshift. The linear correlation
coefficient between the two quantities is -0.094, thus any significant
dependence on the redshift can be excluded. The sample was then
used to evaluate the weighted mean of the spectral index, which is
equal to 1.90, with a dispersion equal to $\pm$0.22. This justifies the
adoption, which follows, of the fixed value $\Gamma$=1.9 for sources with poorer
photon statistics.

The second subset (S2) comprises objects with known z but fewer
counts, hence with spectral fits now performed with $\Gamma$ held
fixed at 1.9.  It includes spectra that, with the best fit
normalization mos/pn frozen before error computing, have
4$\leq$dof$\leq$8.  For these spectra the 90{\%} confidence interval
on N$_H$ was computed with $\Delta \chi^2$ = 2.7. The results for the
36 S2 sources are presented in Table 2, with the same structure as
Table 1. We note that the source 03120008, with 22 dof, belongs
here, because it is extended; however, the fit performed with the 
thermal model {\sc mekal} in XSPEC, usually applied to galaxy clusters, 
is hardly acceptable, and it becomes so only after the
addition of the simple model adopted for the unresolved sources: the
parameters reported are those of the AGN immersed in the extended
source.

\begin{table*}
\begin{center}
\caption{\bf Spectral fits of the subset S2}
\begin{tabular}{l|c|c|c|c|c|c|c|c|c}
\hline
source ID &  type  &  z   &   inst&  N$_H$$^a$ & $\Gamma$ & $\chi^2$/dof &  
F$^b$ &  F$_u$$^b$  & logL$_{2-10keV}$$^c$\\
\hline
03120008 &  ETG  &   0.052  &   pm  & 0.15$\pm_{0.15}^{0.45}$  & 1.9 & 23.2/22 & 2.3 &
2.4  & 41.19\\
03120011  &   AGN1 &     0.753 &   pm  & $<$0.24 & 1.9 & 6.3/8 & 1.7&
1.7  & 43.62\\
03120022  &   AGN1 &     2.140 &   p   & 4.61$\pm_{3.60}^{5.85}$ & 1.9
& 9.1/6 & 3.7 & 3.8 & 45.06\\
03120024  &   AGN1 &     1.838 &   m  & $<$3.36 &
1.9& 1.9/5 & 1.2 & 1.2 & 44.41\\
03120034  &   AGN2 &     0.265 &   pm & 4.33$\pm_{1.35}^{1.81}$ & 1.9
& 10.4/6  & 13.2 & 16.2 & 43.54\\
03120066  &   AGN1 &     1.449  &  pm & $<$1.76 & 1.9 & 9.3/6 & 1.1 & 1.1 & 44.12 \\
0312089a  &   ELG &     0.809 &   m  &  3.05$\pm_{2.89}^{7.34}$ & 1.9
& 9.4/7 & 2.2 & 2.4 & 43.85\\
26900006  &   AGN1  &  0.964 &   m   & $<$0.09 & 1.9 & 5.3/8 & 14.5 & 14.5  & 44.81 \\
26900009  &   AGN1 &   0.995  &  pm  & 0.14$\pm_{0.14}^{0.43}$ &
1.9 & 4.6/8 & 2.0 & 2.0 & 43.98\\
26900016  &   AGN1 &   1.314 &  m  & 0.34$\pm_{0.34}^{0.72}$ & 1.9 &
2.3/6 & 3.7 & 3.7 & 44.54\\
26900022  &   AGN2 &   0.592 &   pm  & 1.05$\pm_{0.50}^{0.69}$ & 1.9 &
4.3/6 & 2.6 & 2.7 & 43.57\\
26900028  &   AGN1  &  0.738 &   pm   &  $<$0.33   &  1.9 &  9.0/5 &
2.4 & 2.4 & 43.75 \\
26000038 & ELG & 0.904 & pm& 4.96$\pm_{1.96}^{3.42}$ & 1.9 & 1.9/5 &
3.5 & 3.9  & 43.9 \\
26900039  &   AGN1  &  0.930 &   pm  &  6.35$\pm_{2.54}^{3.95}$ & 1.9
& 6.3/6 & 6.5 & 7.2 & 44.47 \\
05370019  &   AGN1  &  1.330 &    pm & $<$0.44 & 1.9 & 4.5/8 & 1.5 & 1.5  & 44.16\\
05370020  &   AGN1  &  0.763  &   m  &  $<$0.11 & 1.9 & 13.5/8 & 3.0 &
3.0  & 43.88\\
05370031  &   AGN1  &  3.276  &   pm  & 0.12$\pm_{0.12}^{2.96}$ & 1.9
& 0.3/5 & 1.5 & 1.5 & 45.09\\
05370036  &   AGN1  &  1.329  &   m  &  $<$0.39 & 1.9 & 7.0/8 & 2.7 & 2.7 & 44.42 \\
05370040  &   AGN1  &  1.485  &   pm  & 0.28$\pm_{0.28}^{1.54}$ & 1.9
&1.0/5 & 0.9 & 0.9 & 44.08\\
05370041  &   AGN1  &  1.644 &    pm  & $<$0.79 & 1.9 & 11.5/7 & 0.8 &
0.8  & 44.1\\
05370043  &   AGN2  &  1.797 &    pm  & 10.5$\pm_{4.8}^{9.4}$ & 1.9
& 4.6/8 & 3.1 & 3.4 &  44.83\\
05370123  &   AGN2$^{*}$  &  1.153   &  m   & 6.63$\pm_{4.06}^{21.58}$ & 1.9
& 7.2/5 & 2.7 & 3.0 & 44.32 \\
05370135  &   AGN2  &  0.484  &   pm  &   1.72$\pm^{ 2.93}_{ 1.44}$  &
1.9 &  0.6/5 & 1.2 & 1.3  & 43.05\\
0537042a  &   AGN1  &  1.945  &   pm  & 0.33$\pm_{0.33}^{1.37}$ & 1.9
& 5.6/7 & 1.5 & 1.5 & 44.56\\
50900001  &   AGN2  &   1.049 &    m  &   $<$1.14  &   1.9 &  2.3/4
&2.1 & 2.1 & 44.06 \\
50900013  &   AGN2 &    1.261 &    m  & 2.52$\pm_{2.15}^{4.58}$ &  1.9
& 6.8/6 & 3.0 & 3.1 & 44.42\\
50900036  &   AGN2 &    0.694  &   m  & $<$0.97 & 1.9 &
7.2/6 & 2.3 & 2.3  & 43.67\\
50900039  &   AGN1 &    0.818  &   m  &  $<$0.83 & 1.9 &
6.2/5 & 1.4 & 1.4 & 43.62\\
50900061  &   ETG &     0.324  &   m  &  0.47$\pm_{0.26}^{0.45}$ & 1.9
& 10.8/7 & 3.7 & 3.8 & 43.11\\
50900067  &   AGN1 &    1.076  &   m  & $<$0.57 & 1.9 & 5.8/7 & 3.4 & 3.4 & 44.3 \\
15800005  &   AGN1  &   1.207  &    m  & $<$0.13 & 1.9 & 8.7/8 & 3.7 &
3.7  & 44.45 \\
15800011  &   AGN1 &    2.069  &   m  &     $<$0.50 &   1.9 &  1.5/4 &
 2.5 & 2.5 &  44.85 \\
15800012  &   AGN2 &    0.233  &   m  &  1.63$\pm_{0.54}^{0.74}$ & 1.9 &
10.6/5 & 6.0 & 6.5  & 43.01 \\
15800013  &   ELG &     1.326  &   m  & 1.92$\pm_{1.13}^{2.12}$ & 1.9
& 6.7/5 & 1.9 & 2.0 & 44.29 \\
15800017  &   AGN1 &    1.946 &    m   & $<$0.61 & 1.9 & 4.1/6 & 2.9 &
2.9 & 44.85\\
15800019   &  AGN2 &    1.957 &    m   & 7.26$\pm_{5.45}^{11.67}$ &
1.9 & 6.7/5 & 2.3 & 2.5 &  44.79\\
\hline
\end{tabular}
\end{center}
$^a$ N$_H$ in source frame, units of 10$^{22}$ cm$^{-2}$;
$^b$ Flux in the 2--10 keV band, F: observed, F$_u$: corrected
for absorption, in units of 10$^{-14}$ erg cm$^{-2}$ s$^{-1}$;
$^c$ Log of the luminosity in units of erg s$^{-1}$.
\end{table*}

The third subset (S3) comprises sources without a spectroscopic
redshift.  In terms of number of dof they are a mix of spectra of the
type in S1 and S2. Following the same fit and error procedures,
to show the redshift effect on the results, the spectra were
attributed two fiducial values of z, equal to 1 and 2.
The results for the 11 S3 sources are presented in Table 3, with
the same structure as the previous tables separately for the two
values of z, except that the flux is reported once for each source
under z=2, because it is very similar to that obtained with z=1; in
the second last column the R magnitude of the optical counterpart is
reported. Given this magnitude, four of them could have been
spectroscopically identified: rather than assigning to them
an average value of z using those identified and with similar
magnitude, since they are already well represented by the latter,
it was decided to leave them out from the sample analysis in the next
sections, hence no attempt was made to assign them a
luminosity. Conversely, for the sources fainter than R=23,
in the last column an X-ray luminosity is given,
that has been evaluated, following Paper IV, as 
will be explained in Sect. 4.2.

\begin{table*}
\begin{center}
\caption{\bf Spectral fits of the subset S3}
\begin{tabular}{l|c|c|c|c|c|c|c|c|c|c|c}
\hline
source ID & inst & N$_H$$^a$ & $\Gamma$  & $\chi^2$/dof & N$_H$$^a$ & $\Gamma$
& $\chi^2$/dof & F$^b$ &F$_u$$^b$ & R$^c$ & logL$_{2-10keV}$$^d$ \\

\hline
& & \multicolumn{3}{c|}{z = 1} & \multicolumn{5}{c|}{z = 2}\\
\hline
03120029 &  pm  &  $<$0.56 & 1.9 & 8.2/8  & $<$1.51 & 1.9 & 8.2/8 &
1.2 & 1.2 & 18.8 & \\
03120031 &  pm  &   1.58$\pm_{0.93}^{1.59}$ & 1.9 & 0.9/6  &
4.38$\pm_{2.60}^{4.76}$ &  1.9 & 0.9/5   & 1.7 & 1.8 & 23.6 & 44.29\\
03120045  & pm  & 3.20$\pm_{1.82}^{3.34}$ & 1.9 & 6.1/8 &
9.29$\pm_{5.46}^{10.76}$ & 1.9 & 6.2/8 &   2.8 & 3.0 & 24.4 & 44.81\\
03120065  & pm  & 1.15$\pm_{1.15}^{0.28}$ & 1.9 & 6.6/9 & 3.22$\pm_{3.22}^{7.10}$& 1.9 & 6.6/9 & 1.6 & 1.7 & $\geq$24 & 43.98 \\
26900014  & pm &  0.25$\pm_{0.25}^{0.64}$ & 2.24$\pm_{0.61}^{0.80}$ &
7.0/9 & 0.69$\pm_{0.69}^{1.73}$ & 2.24$\pm_{0.61}^{0.79}$ & 7.0/9 &
1.4 & 1.4 & 21.6 &  \\
26900075  & pm  & 10.2$\pm_{5.9}^{14.2}$ & 1.9 & 0.3/4 &
32.9$\pm_{20.2}^{44.7}$ & 1.9 & 0.4/4 & 3.3 & 4.0 & 24.6 & 45.05 \\ 
05370010  & pm  &  $<$0.45 & 1.82$\pm_{0.52}^{0.75}$ &
12.3/15 & $<$1.26 &  1.83$\pm_{0.30}^{0.52}$ & 12.2/15 & 2.7 & 2.7 & 22.4 & \\
05370012 &  pm  & 0.03$\pm_{0.03}^{0.42}$ & 1.83$\pm_{0.31}^{0.52}$ &
15.9/14 & $<$1.24 & 1.83$\pm_{0.30}^{0.52}$ & 15.9/14
& 2.4 & 2.4 & 22.5 & \\
05370022 &  pm &  0.18$\pm_{0.18}^{0.38}$ & 1.9 & 7.5/6  & 0.51$\pm_{0.51}^{1.04}$ & 1.9 & 7.5/6 & 2.8 & 2.8 & $\geq$23.0 & 44.19\\
05370054 &  pm  & 0.78$\pm_{0.78}^{5.69}$ & 1.69$\pm_{0.80}^{1.92}$ &
4.6/8 &
1.68$\pm_{1.68}^{14.60}$ & 1.61$\pm_{0.71}^{1.73}$ & 4.7/8 & 2.1 & 2.1
& 25.0 & 44.88\\
05370111  & pm  &  7.04$\pm_{4.12}^{9.66}$ & 1.9 & 3.3/8 & 20.1$\pm_{12.2}^{26.5}$ &
1.9 & 3.8/8 & 2.1& 2.3 & 24.5 & 44.69\\
\hline

\end{tabular}
\end{center}
$^a$ N$_H$ in source frame, z=1 or z=2, units of 10$^{22}$ cm$^{-2}$;
$^b$ Flux in the 2--10 keV band, F: observed, F$_u$: corrected
for absorption, in units of 10$^{-14}$ erg cm$^{-2}$ s$^{-1}$;
$^c$ R magnitude of optical counterpart;
$^d$ Log of the luminosity in units of erg s$^{-1}$, only for
sources with R greater than 23, for the redshift values given
in Table 7 (see Sect. 4.2).
\end{table*}

The remaining spectra were treated with the C statistic, and the
results are separately presented in Table 4 for the subset S4 (13
sources with known z, same structure as Table 2), and in Table 5 for the
subset S5 (13 sources with unknown z, same structure as Table 3).  In
S4 there is one source for which N$_H$ is unconstrained (likely due
to spectral complexity combined with poor count statistics); it will
therefore be excluded from the statistical considerations. 
In S5 five sources have their optical counterparts
brighter than R=23, and will be treated like the similar sources in S3, as 
explained above.

 \begin{table*}
\begin{center}
\caption{\bf Spectral fits of the subset S4}
\begin{tabular}{l|c|c|c|c|c|c|c|c|c}
\hline
source ID &  type  &  z   &   inst&  N$_H$$^a$ & $\Gamma$ & Cst/bins
&  F$^b$ &  F$_u$$^b$  & logL$_{2-10keV}$$^c$ \\
\hline
03120016 & AGN2$^{*}$ & 0.841 & pm & 35.5$\pm_{21.6}^{43.1}$ & 1.9 & 22.6/24
& 3.1 & 5.3 & 44.23\\
03120035 & AGN1 & 1.272 & m & 0.23$\pm_{0.23}^{2.15}$ & 1.9 & 19.3/20
& 1.4 & 1.4  & 44.09\\
03120127 & AGN1 & 2.251 & pm & $<$115 & 1.9 & 15.6/18 & 3.4 & 3.4 & 45.07\\
03120181 & ELG & 0.709 & pm & 24.6$\pm_{20.3}^{40.2}$ & 1.9 & 14.0/14
& 1.2 & 2.0  & 43.63\\
03120501 & ETG & 0.205 & m & uncons & 1.9 & 16.5/20 & 1.3 & 1.3 & 42.18 \\
26900072 & ELG & 1.389 & p & 59.3$\pm_{49.5}^{77.7}$ & 1.9 & 14.9/15 &
8.2 & 13.4 & 45.16 \\
05370035 &  AGN1 & 0.897 & p  & 0.18$\pm_{0.18}^{0.85}$ & 1.9 &
13.3/11 & 0.9 & 0.9 & 43.55\\
05370078 & AGN1 & 1.622 & m & 14.2$\pm_{10.0}^{48.4}$ & 1.9 & 24.5/23
& 2.0 & 2.3 & 44.56\\
05370175 & AGN1 & 1.246 & pm & 50.7$\pm_{33.3}^{60.3}$ & 1.9 & 21.9/22
& 2.5 & 4.1 & 44.53\\
0537052a & AGN1 & 1.665 & pm & 0.83$\pm_{0.83}^{1.36}$ & 1.9 & 31.6/20
& 0.8 & 0.8 & 44.12\\
15800025 & ELG & 0.470 & m & 0.29$\pm_{0.28}^{0.61}$ & 1.9 & 16.3/15 &
1.9 & 1.9 & 43.18\\
15800062 & AGN2 & 1.568 & m & 26.3$\pm_{18.1}^{44.7}$ & 1.9 & 25.2/18
& 2.8 & 3.4 & 44.69\\
15800092 & ELG & 0.993 & m & 16.8$\pm_{9.0}^{16.0}$ & 1.9 & 28.6/20 &
3.3 & 4.2  &  44.3\\
\hline
\end{tabular}
\end{center}
$^a$ N$_H$ in source frame, units of 10$^{22}$ cm$^{-2}$;
$^b$ Flux in the 2--10 keV band, F: observed, F$_u$: corrected
for absorption, in units of 10$^{-14}$ erg cm$^{-2}$ s$^{-1}$;
$^c$ Log of the luminosity in units of erg s$^{-1}$.
\end{table*}

\begin{table*}
\begin{center}
\caption{\bf Spectral fits of the subset S5}
\begin{tabular}{l|c|c|c|c|c|c|c|c|c|c|c}
\hline
source ID & inst & N$_H$$^a$ & $\Gamma$  & Cst/bins & N$_H$$^a$ & $\Gamma$
& Cst/bins& F$^b$ &F$_u$$^b$ & R$^c$ & logL$_{2-10keV}$$^d$\\
\hline
& & \multicolumn{3}{c|}{z = 1} & \multicolumn{5}{c|}{z = 2}\\
\hline
03120036  & p  & 1.15$\pm_{0.98}^{1.35}$ & 1.9 & 12.3/18  &
3.22$\pm_{2.72}^{3.93}$ & 1.9 & 13.1/18  & 1.9 & 1.9 & 24.6 & 44.79\\
03120124 & m  & 5.26$\pm_{3.27}^{5.54}$ &1.9 & 11.0/17 &
15.8$\pm_{10.1}^{18.6}$ & 1.9 & 11.1/17  & 2.2 & 2.4 & 22.5  &  \\
26900029 & pm  & 0.75$\pm_{0.57}^{0.90}$ &1.9 & 14.2/19 &
2.03$\pm_{1.53}^{2.60}$ & 1.9 & 14.3/19  & 2.8 & 2.8 & 25.1  & 45.11\\
05370037 & p  & 1.37$\pm_{0.93}^{1.82}$ &1.9 & 17.9/19 &
4.09$\pm_{2.83}^{4.77}$ & 1.9 & 17.5/19  & 4.4 & 4.5 & 21.5  & \\
05370060 & pm  & 0.90$\pm_{0.90}^{1.93}$ &1.9 & 18.5/23 &
2.41$\pm_{2.41}^{5.45}$ & 1.9 & 18.5/23  & 1.0 & 1.0 & 23.9  & 44.11\\
05370072 & pm  & 5.65$\pm_{3.38}^{6.01}$ &1.9 & 15.6/25 &
16.9$\pm_{10.4}^{19.1}$ & 1.9 & 15.6/25  & 1.0 & 1.1 & $\geq$24 & 44.16 \\
05370091 & m  & 24.7$\pm_{19.4}^{54.9}$ &1.9 & 24.6/22 &
55.5$\pm_{40.3}^{98.4}$ & 1.9 & 25.0/22  & 4.2 & 5.7 &  23.7 & 44.26\\
0537011b & m  & 34.0$\pm_{24.9}^{61.6}$ &1.9 & 7.7/11 &
66.4$\pm_{46.3}^{86.2}$ & 1.9 & 9.2/11  & 1.4 & 2.0 & 21.7 & \\
05370153 & pm  & 10.3$\pm_{7.1}^{16.0}$ &1.9 & 25.2/24 &
34.2$\pm_{23.9}^{48.0}$ & 1.9 & 25.2/24  & 1.2 & 1.4 & $\geq$24.6  &
44.54\\
05370157 & pm  & 11.4$\pm_{5.3}^{9.0}$ &1.9 & 27.3/26 &
33.0$\pm_{15.9}^{24.9}$ & 1.9 & 28.5/26  & 1.4 & 1.8 & $\geq$24.5  & 44.52 \\
05370162 & pm  & 5.26$\pm_{4.00}^{11.00}$ &1.9 & 32.5/20 &
15.4$\pm_{12.2}^{36.7}$ & 1.9 & 32.7/20  & 1.3 & 1.5 & 21.6 & \\
0537042b & m  & 8.32$\pm_{4.63}^{18.90}$ &1.9 & 15.6/13 &
22.8$\pm_{12.8}^{36.4}$ & 1.9 & 16.5/13  & 2.2 & 2.5 & 21.5 & \\
0537052b & m  & 7.13$\pm_{3.93}^{6.35}$ &1.9 & 11.7/14 &
22.8$\pm_{13.0}^{20.9}$ & 1.9 & 11.4/14   & 1.7 & 2.0 &  23.7  & 44.34\\
\hline
\end{tabular}
\end{center}
$^a$ N$_H$ in source frame, z=1 or z=2, units of 10$^{22}$ cm$^{-2}$;
$^b$ Flux in the 2--10 keV band, F: observed, F$_u$: corrected
for absorption, in units of 10$^{-14}$ erg cm$^{-2}$ s$^{-1}$;
$^c$ R magnitude of optical counterpart;
$^d$ Log of the luminosity in units of erg s$^{-1}$, only for
sources with R greater than 23, for the redshift values given
in Table 7 (see Sect. 4.2). 
\end{table*}

\begin{figure}
\centering
\includegraphics[angle=0,width=9cm]{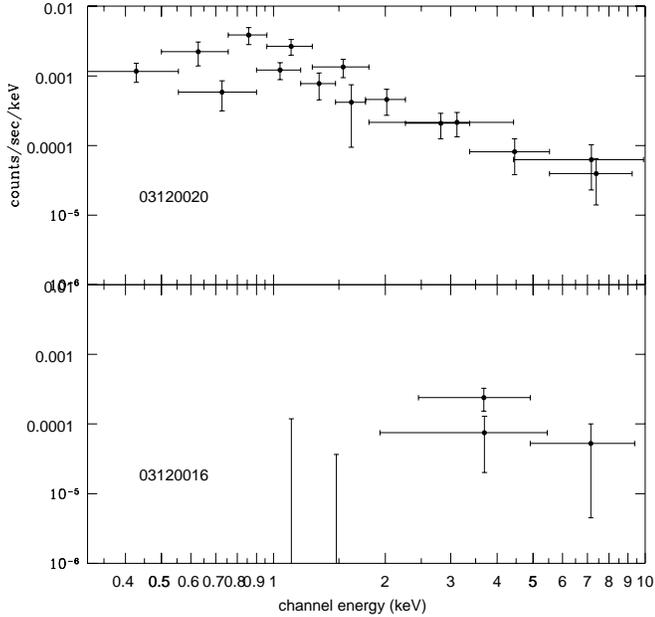}
\caption{The X-ray spectral count distributions of two sources 
with comparable 2--10 keV fluxes.
At the top the unabsorbed source 03120020, at the bottom
the source 03120016, which is evidently affected by strong
absorption.
}
\label{FigAbsSp}%
\end{figure}

Note that, despite the lower number of counts, in
S4 and S5 the 2-10 keV fluxes observed, as estimated from the images 
and reported in Paper IV, are in the same range of values as in S2 and S3.  
This can be due either to larger off-axis angles or to a higher
incidence of large N$_H$ values. The distribution of the off-axis angles
in the S2+S3 sample is however fully consistent with that in the S4+S5
sample (probability of 50\% using the Kolmogorov-Smirnov test). On the
other hand the distributions of the best fit logN$_H$ differ with a
probability higher than 99.98\% and the median best fit logN$_H$ with
their interquartile ranges are 21.5$\pm$1.8 and 23.2$\pm$0.6 for the
S2+S3 and S4+S5 samples respectively.  As an example, Fig. 
\ref{FigAbsSp} shows the spectral counts of two sources in the field of
PKS0312-77 at similar off-axis angles: it is clear that the difference
in the total number of counts is due to a substantial difference in
N$_H$.

In Appendix 1 and 2, a comparison is drawn between the values
of the flux (estimated directly from the images) and 
of N$_H$ (estimated from a count hardness ratio) adopted in
Paper IV, and the same values obtained from the spectral fits. 

\section{A synthesis of the results}

\subsection{The N$_H$ distribution as a function of the flux}

\begin{figure}
\centering
\includegraphics[angle=0,width=9cm]{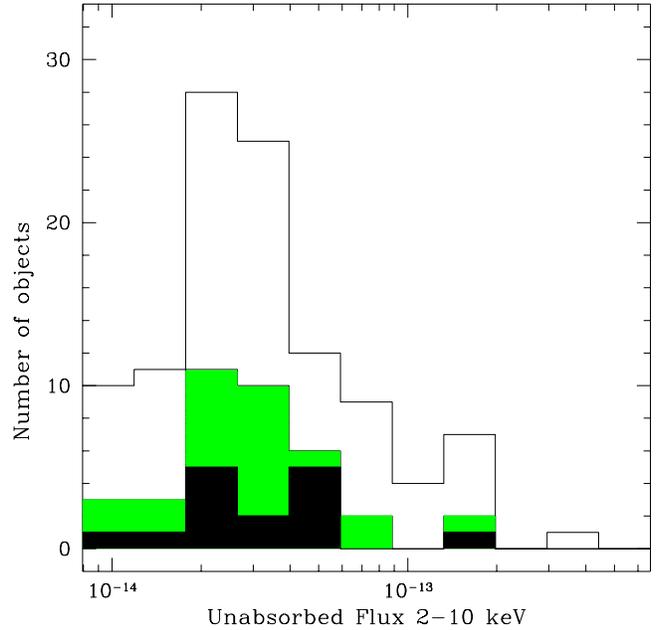}
\caption{The distribution as a function of F$_u$ of the sources in
the sample. See text for the sources excluded from this plot.
In black the sources with N$_H$$>$10$^{23}$ cm$^{-2}$, in gray
those with N$_H$ between 10$^{22}$ and 10$^{23}$ cm$^{-2}$,
in white those with N$_H$$<$10$^{22}$ cm$^{-2}$.
}
\label{FigFuHyst}%
\end{figure}

After the exclusions motivated in the previous section, the
sample is now reduced to 107 sources. From their spectral fits, they
can be subdivided into three
categories, according to the best fit value of N$_H$: those with
N$_H$$<$10$^{22}$ cm$^{-2}$, those with N$_H$ between
10$^{22}$ and 10$^{23}$ cm$^{-2}$, those with N$_H$$>$10$^{23}$
cm$^{-2}$. For the sources without redshift there is of course a
difference according to whether z= 1 or 2 is adopted: in this case
the source numbers in each category were first obtained separately with
z=1 and with z=2, then their mean value was used. Fig.
\ref{FigFuHyst} shows a hystogram of the F$_u$ distribution
of the 107 sources, where those falling into each of the three
categories are indicated.

From Fig. \ref{FigFuHyst} it can be immediately appreciated that, with the
exception of the brightest one, 106 sources have fluxes between
0.8$\times$10$^{-14}$ and 20$\times$10$^{-14}$ erg cm$^{-2}$ s$^{-1}$. This sample can be
subdivided into two flux intervals, defined as follows.  In F$_{u1}$
(from 0.8$\times$10$^{-14}$ to 5$\times$10$^{-14}$ erg cm$^{-2}$ s$^{-1}$) there are 82
sources, with a median flux equal to 2.4$\times$10$^{-14}$ erg cm$^{-2}$ s$^{-1}$; in
F$_{u2}$ (from 5$\times$10$^{-14}$ to 20$\times$10$^{-14}$ erg cm$^{-2}$ s$^{-1}$) there
are 24 sources, with a median flux equal to 10$\times$10$^{-14}$ erg cm$^{-2}$ s$^{-1}$.

The fraction of sources for each of the three categories defined above,
separately for the two flux intervals, is given in Table 6.

\begin{table}[h]
\begin{center}
\caption{\bf Fraction of absorbed sources$^c$}
\begin{tabular}{l|c|c|c}
\hline
N$_H$ (cm$^{-2}$) &$<10^{22}$& $10^{22}-10^{23}$& $>10^{23}$\\
\hline
F$_{u1}$$^a$ & 52.5/82 & 19/82 & 12.5/82\\
&48.5/82 &20.8/82 &12.7/82\\
F$_{u2}$$^b$ & 18/24 & 3/24 & 1/24\\
&17.9/24 &3.1/24 &3.0/24\\
\hline
\end{tabular}
\end{center}
$^a$ F$_{u1}$: interval from 0.8$\times$10$^{-14}$ to 5$\times$10$^{-14}$ erg cm$^{-2}$ s$^{-1}$;
$^b$ F$_{u2}$: interval from 5$\times$10$^{-14}$ to 20$\times$10$^{-14}$ erg cm$^{-2}$ s$^{-1}$;
$^c$ The fractions in the two lines for each flux interval were obtained
as described in the text (Sect. 4.1).
\end{table}

In order to take into account the uncertainties associated to the best fit values of the
absorption columns, a procedure was set up to weigh each source with
the probability that it falls into each of the N$_H$ categories. This
procedure is based on an analytic approximation (which turns out
to resemble closely the combination of two gaussians with different sigmas, one
for the values above, the other for those below the best fit N$_H$) to the probability
distribution of N$_H$, 
estimated with XSPEC, using the {\sc steppar}
command, for a number of objects selected in such a way as to properly
represent the sources in the five subsets. The results differ
only slightly from the previous ones, and are also given in Table 6,
the second line for each of the two flux intervals. In the
following the results obtained in this way will be used.

\subsection{Fraction of absorbed sources as a function of the luminosity}

The spectral fit results can also be used to diagnose whether
the strength of the photoelectric absorption might be a
function of the luminosity. Despite the relatively minor
numerical contribution of the S3 and S5 sources with R
fainter than 23, a higher fraction of them, compared with
the other subsets, is affected by large values of N$_H$, and
cannot therefore be dismissed. To avoid a fictitious concentration
of them around some value of the luminosity if a nominal
fixed value of z were adopted, a stochastical procedure, based on the
ratio between X-ray and optical fluxes (X/O), already adopted 
in Paper IV, was followed to assign 
individual values of z. These values are given in Table 7, and
the luminosities in Tables 3 and 5 were calculated accordingly.
We note that the estimated values of z are roughly in agreement
with the limits photometrically derived, for some of them,
using the R--K colour by Mignoli et al. (2004). In addition,
we point out that Table 6 would not be significantly different
if these values of z had been used in the
spectral fitting.

\begin{table}[h]
\begin{center}
\caption{\bf Estimated values of z for 15 sources with R$>$23} 
\begin{tabular}{|l|c|}
\hline
source ID & z$_{est}$\\
\hline
03120031 &  1.3\\
03120036  &  2.0\\
03120045  & 1.7\\
03120065  & 1.0\\
26900029 &   2.3\\
26900075  & 1.9\\
05370022 &  1.0\\
05370054 & 2.1\\
05370060 &  1.4 \\
05370072 &  1.4 \\
05370091 &  0.8 \\
05370111  &1.7\\
05370153 &  1.8 \\
05370157 &  1.6\\
0537052b & 1.3 \\
\hline

\end{tabular}
\end{center}
\end{table}

\begin{figure}
\centering
\includegraphics[angle=0,width=9cm]{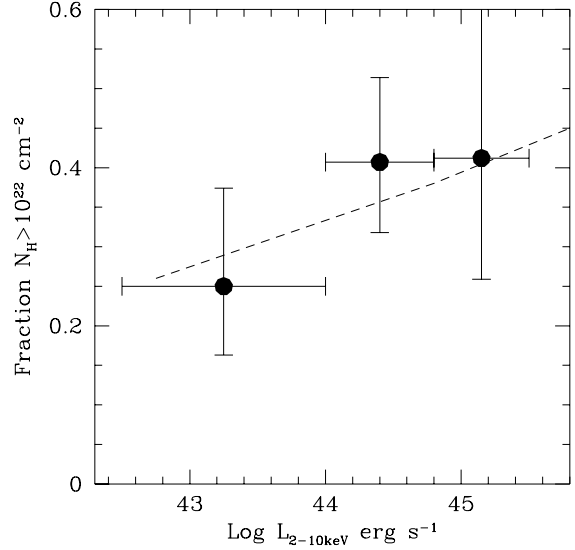}
\caption{The fraction of sources with N$_H$$>$10$^{22}$ cm$^{-2}$
(68\% poissonian errors) as a function of their luminosity. The sources with unknown z
are included, as explained in the text. The broken line represents the
expectation if such a fraction is intrinsically independent of
the luminosity, but see text for comments on this issue in Sect. 4.2 and
Sect. 5.
}
\label{FigAbsLum}%
\end{figure}

The result given in Fig. \ref{FigAbsLum} shows no evidence of a 
luminosity dependence. However, to judge this result properly,
it is necessary to take into account the
selection effects introduced by the inhomogeneity
of the flux limit within any XMM image. The broken line
in the same figure, in very good agreement with the
data points, takes this bias into account, on the basis
of assumptions to be described in Sect. 5, among which there is
one, immediately relevant to the issue, which states
that the fractional distribution of N$_H$ is independent of
the luminosity. It must be stressed, however, that
the latter statement is of restricted value, and need
not remain valid when deeper X-ray
surveys are also taken into consideration (see Ueda et al. 2003).
Such surveys are indeed necessary to expand our knowledge 
at higher redhifts and lower luminosities, in order to better
investigate the incidence of absorption, its higher values
in particular, than could be done within the flux limit of our sample, as
already noted when comparing the subsets S4/S5 to the subsets
S2/S3. This point will be revisited in Sect. 5.

\subsection{N$_H$ versus the ratio X/O and the optical 
spectroscopic classification}

\begin{figure}
\centering
\includegraphics[angle=0,width=9cm]{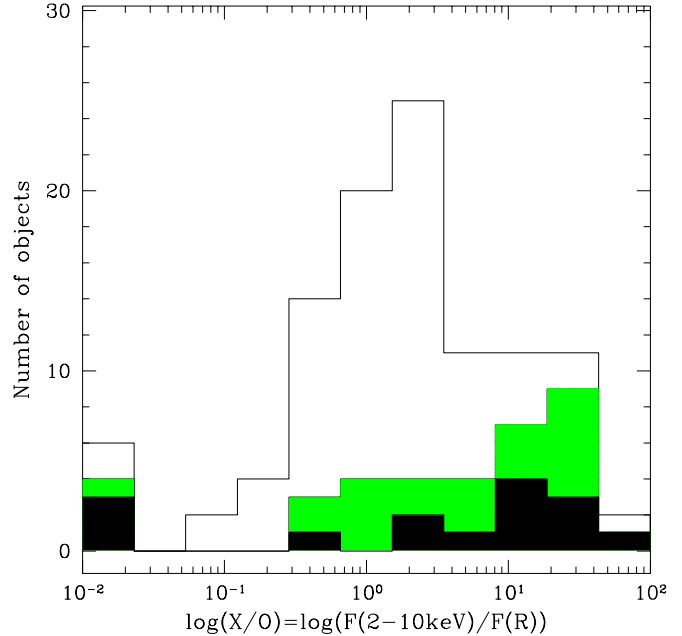}
\caption{The X/O distribution for the sources in
the sample. In black the sources with N$_H$$>$10$^{23}$ cm$^{-2}$, in gray
those with N$_H$ between 10$^{22}$ and 10$^{23}$ cm$^{-2}$,
in white those with N$_H$$<$10$^{22}$ cm$^{-2}$.
}
\label{FigX/O}%
\end{figure}

In Paper IV the existence was emphasized of a close correlation between
the X/O ratio and the X-ray luminosity for objects optically
classified as non--broad--line AGN. Figure \ref{FigX/O} illustrates
the fact that the fraction of highly absorbed sources is greater for
the large than for the small values of X/O, thus confirming the latter
as a fairly reliable diagnostic parameter for a preliminary
classification of high luminosity, high obscuration AGN.  Indeed, 16
out the 57 objects with logL$_{2-10keV}>44$ in the sample of sources
with spectroscopic redshifts (S1+S2+S4) have best fit logN$_H>22$ (11
of these objects have logN$_H>22$ at $>90\%$ confidence level, 7 have
best fit logN$_H>23$). If we consider also the sources without a
spectroscopic redshift (S3+S5) to which a luminosity was assigned as
described in Sect. 4.2, the number of high luminosity sources with
logN$_H>22$ increases to 29 out of 71.  The fraction of high
luminosity AGN (QSO) obscured in X-rays is then at least 28\%, most
likely about 40\%. This fraction can be translated, taking into
account the sky coverage, into a surface density of highly obscured QSOs
of $\sim48$ deg$^{-2}$, at the flux limit of $\sim10^{-14}$ erg cm$^{-2}$ s$^{-1}$ of
the HELLAS2XMM 1df survey. This density should be compared with
the estimate (Mainieri et al. 2002)  of $\sim69$ deg$^{-2}$, 
based on six objects only, in the 0.5--7 keV band at a flux limit 
of 1.6$\times$10$^{-15}$ erg cm$^{-2}$ s$^{-1}$.

There are cases where the optical spectroscopic classification
turns out the opposite of the X-ray classification, when the latter is
based on the X-ray obscuration (see in particular Akiyama et al. 2003,
Brusa et al. 2003).  In the S1, S2 and S4 subsets one finds five
objects (six if 03120127, with its very large upper limit on N$_H$, is
included) optically classified as AGN1, whose best fit logN$_H$ is
greater than 22. In one case, 05370175, the absorbing column is
greater than 10$^{23}$ at $>90\%$ confidence level.  The fraction, 5
or 6 out of 60, is about 10\%, in agreement with the finding (3/29) by
Page et al. (2003). We note that the 6 objects with
logN$_H>22$ all have logL$_{2-10keV}>44$, while the 17 AGN1 with
logL$_{2-10keV}<44$, i.e. the Seyfert 1 objects, have logN$_H<22$:
this difference cannot be attributed to a redshift dependent bias,
however its significance is not high,
according to the Fisher exact probability test (Siegel 1956) it amounts
only to 89\%.

The anomalous cases might reflect the existence of substantial
variance in the dust to gas ratio, or alternatively of a geometrical
separation with respect to the line of sight between the X-ray
absorbing gas and the gas and dust in front of the broad line
region. From a purely empirical side, one should not forget that
variability may also play a role. One example is NGC 4151, which is
characterized by a fairly dense, and variable, X-ray absorbing column;
this object, when it was repeatedly observed with the IUE
satellite, in some epochs simultaneously in the X-rays with EXOSAT,
displayed impressive differences in the width of the permitted lines,
from very broad to very narrow, correlated with changes in brightness,
but without any evident correlation with the amount of N$_H$ (Perola
et al. 1986, Fiore et al. 1990, Ulrich 2000 and references therein).
None of these hypotheses, though, give an obvious answer to the question
why the anomalous cases should appear to be more common
at QSO luminosities.

One also finds four objects classified as AGN2, with logN$_H$ less
than 22. Here the most probable origin of the discrepancy is the
complexity of the X-ray spectra, which is found in detailed studies of
bright sources (e. g. Turner et al. 2000, for the variable case of NGC
7582). As noted in Sect. 3, the simple model fit adopted aims to
obtain an effective value of N$_H$, which for this paper is the 
relevant quantity.

\section{Discussion}

\begin{figure}
\centering
\includegraphics[angle=-90,width=9cm]{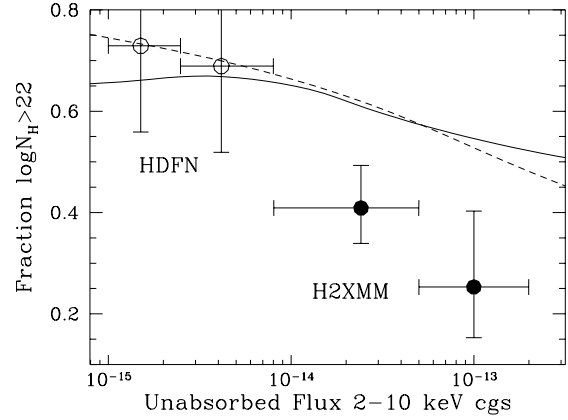}
\caption{The fraction of sources with logN$_H$ greater than 22,
in the two F$_u$ intervals from Table 6 (full circles), compared with the
predictions based on the model (solid line) by Comastri et al. (1995,
see also Comastri et al. 2001), and the model B (dashed line)
by Gilli et al. (2001).
The points in the two lower flux intervals (empty circles) were derived
from the CDFN survey (see text). 
}
\label{FigAbs/Comastri}%
\end{figure}

In Fig. \ref{FigAbs/Comastri} the fractions of objects with logN$_H$
greater than 22, from Table 6, second line for each of the two
flux intervals, are reproduced. The error bars (68\%) are based purely on
poissonian statistics, for an immediate comparison with the
results collected in Piconcelli et al. (2003).  

In the same figure the solid line represents
the prediction of one of the so far most popular XRB synthesis models
(Comastri et al. 1995, see also Comastri et al. 2001 for the N$_H$
distribution), the dashed line model B in Gilli et al. (2001). 
The binomial distribution is used to estimate the significance 
of the discrepancy between the two predictions and the observational
results. This significance turns out equal to 99.999\% for both models. 

This result consolidates the finding by Piconcelli et al. (2003), based on an
XMM--Newton sample, comparable in size and flux coverage with the one
used here (15 sources are in common with the present sample) but with
a much lower percentage of spectroscopic identifications (about 40\%).
It also supports the findings by Mainieri et al. (2002, as derived
from their sample analysis by Piconcelli et al. 2003) with XMM--Newton in a flux interval
similar to that considered in this paper, by Akiyama et al. (2000) with ASCA around
5$\times$10$^{-13}$ erg cm$^{-2}$ s$^{-1}$, and by Caccianiga
et al. (2004) with XMM-Newton around 10$^{-13}$ erg cm$^{-2}$ s$^{-1}$.

For flux values between 1 and 8$\times$10$^{-15}$ erg cm$^{-2}$ s$^{-1}$ Fig.
\ref{FigAbs/Comastri} shows our estimates (again with 68\%
poissonian errors) derived from data on
the Chandra Deep Field North (CDFN) given in Brandt et al. (2001) and
Barger et al. (2002).  The values of N$_H$ were obtained from flux
hardness ratios, thus they are not as reliable as those obtained from
a spectral fit. Taking these estimates at face value, the discrepancy
seems to disappear as one goes below F$_u$=10$^{-14}$ erg cm$^{-2}$ s$^{-1}$.

To investigate the origin of the discrepancy, given the modest
difference in the predictions of the two models, one can concentrate 
for simplicity on the first. To this effect it is useful to recall the
three main assumptions in Comastri et al. (1995), namely: a) the LF is
characterized by a double power law shape and pure luminosity 
evolution (PLE); b) the fractional
distribution of the N$_H$ values is independent of source luminosity
and redshift; c) this distribution is {\it adjusted} to comply with the
spectral shape of the XRB.

\begin{figure}
\centering
\includegraphics[angle=-90,width=9cm]{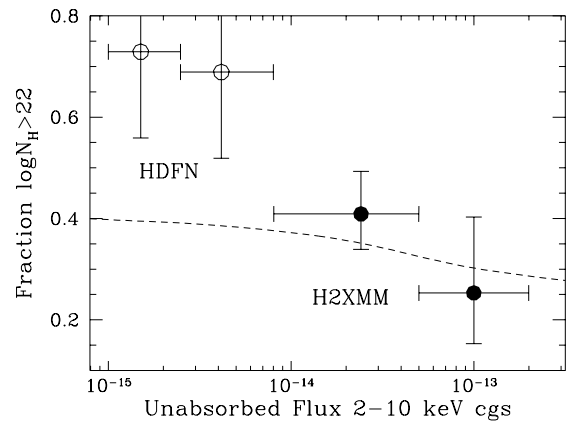}
\caption{The fraction of sources as in Fig. \ref{FigAbs/Comastri}, here compared
with the prediction which is obtained when, in the Comastri et al. (1995, 2001) 
model, the N$_H$ distribution is changed to a flat one, as described in the text.
}
\label{FigAbs/Noi}%
\end{figure}

The broken line in Fig. \ref{FigAbs/Noi} represents the expectation
for the sample of objects used here, when the assumptions a) and b)
are maintained, but, as an excercise, the N$_H$ distribution is adopted,
which corresponds to the broken line in Fig. 4. This distribution
differs substantially from the one given 
in Comastri et al. (2001). The fractional
value, per logN$_H$ decade, is 0.3 between 20 and 21, then it drops to
0.175 and stays constant up to logN$_H$=25, a column
density above which the absorber is Compton thick, to the extent that
the direct emission is practically undetectable between 2 and 10 keV
in the flux range explored.  Correspondingly, Fig. \ref{FigAbs/Noi}
shows a prediction which is radically different from the one in Fig.
\ref{FigAbs/Comastri}, and, not surprisingly after the agreement found
in Fig. \ref{FigAbsLum}, is in reasonably good agreement with the
results from this sample.  Notably though, the CDFN points are now in
excess with respect to the prediction.

The conclusion from this excercise is that, down to a flux level
where only 35\% of the XRB is resolved, in order to better
reproduce the observations a change in the N$_H$ distribution
would be sufficient. It goes almost without saying that the same
excercise (the N$_H$ distribution being different from the one
adopted in Comastri et al. 1995, 2001) fails to reproduce
satisfactorily the spectral shape of the XRB. Thus a more complex
approach is needed, like the one followed by Ueda et al. (2003), which
takes into account simultaneously the LF, its evolution, and the N$_H$
distribution, the latter in principle as a function of luminosity and redshift:
but this approach can only be pursued using a sample encompassing
wider flux and luminosity ranges, as anticipated in Sect. 4.2, and
is being pursued using the full sample adopted in Paper IV (La Franca et
al., in prep.).

\section{Conclusions}

Starting from 117 sources from the HELLAS2XMM 1df survey (Paper IV),
after the exclusion of 1  source with N$_H$ unconstrained and 9 sources with unknown z
and R$<$23, the spectroscopic analysis of the remaining 107 X-ray
spectra (86\% with spectroscopic redshift) led to the following main
result. The fraction of the 106 sources with logN$_H$$>$22 in the flux
interval 0.8--20$\times$10$^{-14}$ erg cm$^{-2}$ s$^{-1}$  is inconsistent, at the
99.999\% confidence level, with the predictions of two well known XRB
synthesis models, one by Comastri et al. (1995), the other by Gilli et
al. (2001, their model B). This result consolidates the discrepancy also
found by other authors in this flux interval, as mentioned in the
previous section.

As an exercise for the Comastri et al. (1995) model, leaving unchanged
all other assumptions, the adoption of a simple and different
intrinsic distribution of the source percentage per decade of N$_H$,
from log$N_H$=20 to 25 (which is consistent with the results from the
present sample, see Fig. 4), leads to a much better agreement down
to 10$^{-14}$ erg cm$^{-2}$ s$^{-1}$, but fails to reproduce 
the much larger percentage
of absorbed sources, derived from the CDFN survey, in the 10$^{-15}$
to 10$^{-14}$ erg cm$^{-2}$ s$^{-1}$  flux interval.

A study encompassing a much wider flux range, with a sufficiently 
large sample of objects (such as the one used in Paper IV), that should
tackle simultaneously the problems of the shape and evolution of the
LF, of the N$_H$ distribution as a function of luminosity and cosmic
epoch and eventually the XRB synthesis (with an approach akin to that
followed by Ueda et al. 2003), goes beyond the scope of the present
paper, and will be the subject of La Franca et al. (in prep.).

An important result, which basically confirms what was found in Paper
IV, is that in our sample at least 28\%, most likely about 40\% of AGN
with logL$_{2-10keV}>44$ (that is of the QSO) are obscured in X-rays
(log$N_H>22$). This fraction can be translated, taking into account
the sky coverage, into a surface density of highly obscured QSO of
$\sim48$ deg$^{-2}$, at the flux limit of $\sim10^{-14}$ erg cm$^{-2}$ s$^{-1}$ 
of the HELLAS2XMM 1df survey.

As a side issue, note that in the sample studied, while a value of the
parameter log(X/O) much greater than unity confirms itself to be
strongly indicative of high obscuration in high luminosity AGN, 
as shown in Paper IV, there are 5 or 6 out of 60, that
is about 10\% of sources, with logN$_H$$>$22, that are optically
classified as AGN1, in agreement with a previous finding by Page et
al. (2003; see also Brusa et al. 2003, Akiyama et al. 2003). Notably
they are all concentrated at logL$_{2-10keV}>44$. Among various
possibilities, it is pointed out that variability may be one of the
causes of this inconsistency.

\section{Appendix 1}

\begin{figure}
\centering
\includegraphics[angle=0,width=9cm]{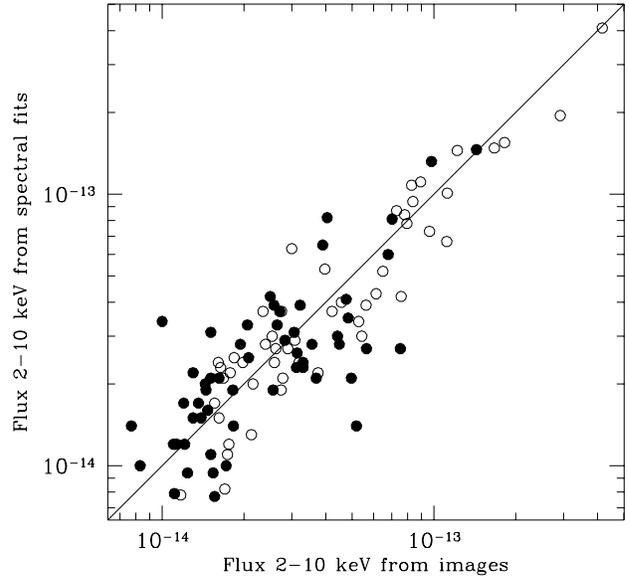}
\caption{Comparison of the best fit flux with that obtained
from the images and used in Paper IV. Open circles are objects
with $N_H<10^{22}$ cm$^{-2}$, filled circles are objects with
$N_H>10^{22}$ cm$^{-2}$. The solid line represents a one to one
relationship.
}
\label{FigFluxfit/fluxim}%
\end{figure}

In PaperIV the source fluxes were derived directly from the counts
image, with a conversion factor appropriate to the filter in front of
the camera and to a spectral shape with $\Gamma$=1.8. In Fig.
\ref{FigFluxfit/fluxim} these fluxes are compared with those obtained
from the detailed spectral fits presented here, which are
more accurate.  The correlation is evidently good, and the
points are distributed uniformly around a one to one relationship; no
large systematic deviation in one sense or the other occurs when
computing fluxes from the images, using the recipe of Baldi et
al. (2002). This applies to both unobscured and obscured sources.  On
the other hand, there is a scatter which obviously increases toward
low fluxes. The standard deviation of the ratio between the two flux
estimates is 50\% and 30\% in the F$_1$ and F$_2$ flux ranges
respectively.

\section{Appendix 2}

\begin{figure}
\centering
\includegraphics[angle=0,width=9cm]{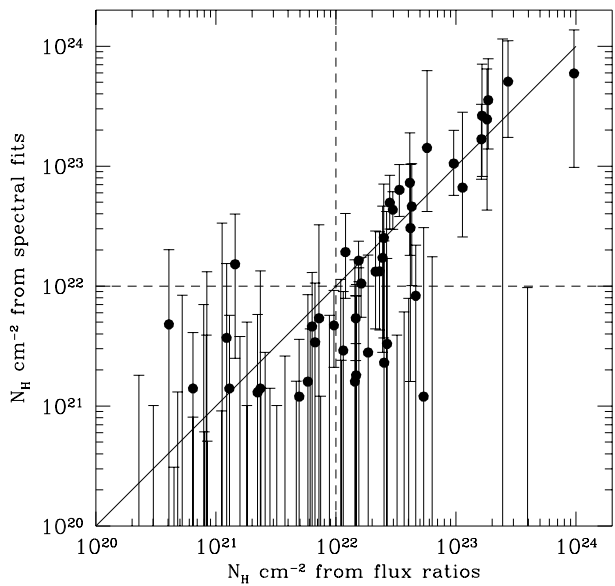}
\caption{Comparison of the best fit values of N$_H$ (with 90\% error
bars) to those estimated with
the Softness Ratio technique used in Paper IV. The solid line represents
a one to one relationship, the two dashed lines divide the figure in
four quadrants with log$N_H$ higher or lower than 22.
}
\label{FigNhfit/NhHR}%
\end{figure}

In Paper IV the values of N$_H$ individually used to correct the
observed flux, and hence the luminosity, for absorption were obtained
by means of a count Softness Ratio (SR). These values are given in
Fig. 9 against those obtained from the spectral fits (for the
sources with a spectroscopic redshift).  As for the flux values in
Fig. 8, a rather satisfactory correlation is present with the values
obtained from the spectral fits.  The obvious limitation in the SR
technique is that the error estimate is less reliable, but in a
statistical sense the results obtained with this simple approach are
sufficiently representative of the sample properties.  Nevertheless,
Fig. \ref{FigNhfit/NhHR} suggests the presence of a systematic error
affecting one of the two $N_H$ estimates (very likely that obtained
from the SR). The number of objects in the lower-right quadrant is
significantly higher than that in the upper-left quadrant, that is the
SR technique provided a slighly higher number of objects with a
``nominal'' 22$<$log$N_H<$22.5. A nice correlation is recovered for
higher values of the absorbing column, which are of course easier to
detect. From this experience one might conclude, as a cautionary
remark, that the SR technique tends to over-estimate the true value 
of the absorbing column around logN$_H$=22 by $\approx0.3$.

\section{Acknowledgments}
Financial support is acknowledged from the italian MIUR (Cofin-03-2-23),
INAF (grant 270/3003) and ASI (grant I/R/057/02).

\end{document}